\documentclass{article}
\usepackage{amsfonts}

\usepackage{graphicx}
\usepackage{amsmath}
\usepackage{color}

\begin{document}

\title{Semigroup of positive maps for qudit states and entanglement in tomographic
probability representation}
\author{V.I. Man'ko$^a$, G. Marmo$^b$, A. Simoni$^b$, F. Ventriglia$^b$ \\
{\footnotesize \textit{$^a$P.N.Lebedev Physical Institute, Leninskii
Prospect 53, Moscow 119991, Russia }}\\
{\footnotesize {(e-mail: \texttt{manko@na.infn.it})}}\\
\textsl{{\footnotesize {$^b$Dipartimento di Scienze Fisiche dell'
Universit\`{a} ``Federico II" e Sezione INFN di Napoli,}}}\\
\textsl{{\footnotesize {Complesso Universitario di Monte S. Angelo, via
Cintia, 80126 Naples, Italy}}}\\
{\footnotesize {(e-mail: \texttt{marmo@na.infn.it, simoni@na.infn.it,
ventriglia@na.infn.it})}}}
\maketitle

\begin{abstract}
Stochastic and bistochastic matrices providing positive maps for
spin states (for qudits) are shown to form semigroups with dense
intersection with the Lie groups $IGL(n, \mathbb{R})$ and $GL(n,
\mathbb{R})$ respectively. The density matrix of a qudit state is
shown to be described by a spin tomogram determined by an orbit of
the bistochastic semigroup acting on a simplex. A class of
positive maps acting transitively on quantum states is introduced
by relating stochastic and quantum stochastic maps in the
tomographic setting. Finally, the entangled states of two qubits
and Bell inequalities are given in the framework of the
tomographic probability representation using the stochastic
semigroup properties.\newline \noindent\textit{Key words} Quantum
tomograms, semigroups, positive maps, entangled states.\newline
\end{abstract}

\section{ Introduction}

The description of a physical system admitting a probabilistic
interpretation, be it classical or quantum, requires two
collections of objects called states and observables, say
$\mathcal{S}$ and $\mathcal{O}$\ respectively, along with a
pairing $\mu $ associating with any state $\rho $ and observable
$A$ a Borel probability measure on the real line $\mathbb{R}.$ If
$A$ is measured while the system is in a state $\rho ,\mu _{A,\rho
}$ represents the probability distribution for the observed values
of $A.$ Thus if $E\subseteq \mathbb{R}$ is a Borel set, $\mu
_{A,\rho }(E)\in \lbrack
0,1] $ is the probability that the measured value of $A$ will be in the set $%
E$ when the system is known to be in the state $\rho .$ From a
general point of view, above properties seem to be the minimal
features that any physical system should possess.

This approach has been studied by several authors, for instance
one can find a nice discussion by G. Mackey\cite{Mackey}. The set
$\mathcal{S}$ describes the basic mathematical structure we are
dealing with, while $\mu _{A,\rho }$ provides us with a physical
interpretation. This probabilistic point of view
is compatible with convex combinations on the space of states, indeed if $%
\rho _{1}$ and $\rho _{2}$\ give rise to probability
distributions, by setting $\mu _{A,\lambda \rho _{1}+(1-\lambda
)\rho _{2}}=\lambda \mu _{A,\rho _{1}}+(1-\lambda )\mu _{A,\rho
_{2}}$ we define a new probability distribution when $0\leq
\lambda \leq 1$. Usually one requires some additional structure
telling us how the system changes from time $s$ to a
later time $t,$ i.e. requires the existence of a family of mappings $U_{t,s}:%
\mathcal{S\rightarrow S}$ representing the dynamics and called
evolution operator. The requirement that a state at a given time
determines the state at a later time forces us to postulate the
semigroup property
\begin{equation}
U_{t_{2},t_{1}}=U_{t_{2},s}\circ U_{s,t_{1}}
\end{equation}
with $U_{t,t}$ the identity. Within this setting
\begin{equation}
\mu :\mathcal{S}\times \mathcal{O}\rightarrow \{\mathrm{Borel\
probability\ measures\ on\ }\mathbb{R}\}
\end{equation}

A subset of observables is said to be a tomographic set $\tau $ if
it allows to identify the state $\rho $ (to ``reconstruct''\ the
state) when $\left\{ \mu _{A,\rho }\right\} _{A\in \tau }$ is
known. Very often $\tau $\ is
generated by acting with a group $\mathcal{G}$ on some fiducial observable $%
A_{0},$ i.e. it is the orbit of $\mathcal{G}$\ in $\mathcal{O}$ through $%
A_{0}.$ According to the group we use and the fiducial observable
we start with, we deal with symplectic tomography, photon-number
tomography and so on. While this approach is general enough to
allow us to deal both with classical and quantum systems, here, to
be more definite, we shall consider a quantum system with a finite
number of levels.

States for quantum systems with a finite number of levels will be
thought of as the spin states (or qudits) they can be described by
density matrices which are hermitian nonnegative $(2j+1)\times
(2j+1)$ matrices with unit trace. The linear maps of the spin
states, positive maps, can be described by $(2j+1)^{2}\times
(2j+1)^{2}$ matrices with special properties\cite {Sudar61}.
Recently it was shown\cite{Dod,JETPOlga,OlgaBeppePhysScr} that
qudit states can be described by probability distributions of
random spin projection (called tomogram) depending on the
direction of the quantization axis. In view of this the geometry
of qudit states can be associated with the geometry of a simplex
and the set of positive maps of qudit states can be associated
with stochastic and bistochastic matrices moving points on the
simplex. The aim of this work is to find the connection of spin
tomograms with a unitary matrix containing eigenvectors of the
density matrix of a qudit state and a point on the simplex which
has the eigenvalues of the density matrix as its coordinates.

Another aim of the work is to define positive maps of qudit states
through the transitive actions of both the unitary group on the
eigenvectors of the density matrix and the stochastic matrix
semigroup on the eigenvalues of the density matrix regarded as
points of the simplex.

The qudit states of multipartite systems can be either separable
or entangled. We formulate the properties of a qudit tomogram,
which is the joint probability distribution of two spin
projections on their own quantization axes, able to distinguish
separable and entangled states. We consider the Bell
inequalities\cite{Bell64,CHSH} in the context of the properties of
stochastic matrices constructed by using spin tomograms. The
Cirelson\cite{Cirel} bound $2\sqrt{2}$ for the Bell-CHSH
inequality of two qubits will be connected with some properties of
a universal stochastic matrix obtained from the tomographic
probability distribution describing maximally entangled two
spin-1/2 states. The connection of positive maps with the
semigroup of stochastic matrices provides the possibility to find
a new relation of the maps
with the Lie group of the general linear real transformations $GL(n,\mathbb{%
R})$ for bistochastic matrices and with the inhomogeneous group $%
IGL(n,\mathbb{R})$ for stochastic matrices.

This connection (which seems to have been unknown) provides a
possibility to construct unitary representations of stochastic and
bistochastic semigroups by reducing known infinite dimensional
unitary irreducible representations of the Lie groups to the
subsets of the Lie groups which are the semigroups under
consideration.

The paper is organized as follows: in section 2 we review the spin
tomography approach for one and two qudits. Examples of a qutrit
 and two qubit states in tomographic probability representation are
studied in section 3. The relation of stochastic and bistochastic
semigroups with Lie groups is discussed in section 4, mainly in
the case a qutrit. In section 5 a class of positive maps acting
transitively on quantum states is introduced by relating
stochastic and quantum stochastic maps in the tomographic setting.
The relation of stochastic matrices with Bell inequality violation
for entangled states of two qubits is discussed in section 6. Some
conclusions and perspectives are finally drawn in section 7.

\section{Spin tomograms and unitary group}

As it was shown in \cite{Dod,Olga,SudBeppe} the qudit state
described by a $(2j+1)\times (2j+1)-$matrix $\rho $ can be also
described by a tomographic probability distribution function, or
tomogram, $\mathcal{W}(m,U)\geq 0$ where $m$ is the spin
projection: $ m=-j,-j+1,\ldots ,j-1,j;$ and $U$ is a unitary
$(2j+1)\times (2j+1)-$matrix. This matrix can be considered as a
matrix of an irreducible representation of the rotation group
depending on two Euler angles $\phi ,\theta $ determining the
direction of quantization (or a point on the Bloch sphere
$S^{2}$). The physical meaning of the tomogram $\mathcal{W}(m,U)$\
is that, in the spin state with the given density matrix $\rho ,$
it gives the probability to
obtain $m$ as spin projection on the direction determined by the two angles $%
\phi ,\theta $. It corresponds to choose $\left\{ U^{\dagger
}J_{z}U\right\} $ as tomographic set of isospectral observables,
where $J_{z}=\sum_{m=-j}^{j}m\left\vert m\right\rangle
\left\langle m\right\vert $ is one of the generators of the
irreducible representation of the rotation group, so that
$\mathcal{W}(m,U)$ is nothing but the value of the concentrated
measure $\mu _{U^{\dagger }J_{z}U,\rho }$ at the spectral point
$m:$
\begin{equation}
\mathcal{W}(m,U)=\mu _{U^{\dagger }J_{z}U,\rho
}(m)=\mathrm{Tr}U^{\dagger }\left\vert m\right\rangle \left\langle
m\right\vert U\rho =\left\langle m\left\vert U\rho U^{\dagger
}\right\vert m\right\rangle . \label{tomdef}
\end{equation}
The probability distribution is obviously nonnegative and
normalized, i.e.
\begin{equation}
\sum\limits_{m=-j}^{j}\mathcal{W}(m,U)=1  \label{normcond}
\end{equation}
for any direction of the quantization axis. The spin tomogram can
be also regarded as the diagonal matrix element of the rotated
density matrix $U\rho U^{\dagger }$ in the natural basis
$\left\vert m\right\rangle .$

The relation is invertible and knowing the tomogram
$\mathcal{W}(m,U)$\ for the matrices $U(\phi ,\theta )$ of an
irreducible representation of $SU(2)$ one obtains the density
matrix $\rho $ by means of a linear transform\cite{Olga,SudBeppe}
which is the analog of the integral Radon transform but in the
space of qudit states. Thus the quantum state of a a qudit (a
spin-$j$ state) is known if the probability distribution
$\mathcal{W}(m,U)$ of random
spin projection as a function of the unitary matrix $U$ is known. The tomogram $%
\mathcal{W}(m,U)$ can be used, consequently, in alternative to
spinors (wave functions) or density matrices for describing spin
states. The information on the spin state contained in the
tomogram is redundant since it is sufficient to know the tomogram
only for several directions determined by a set of angles $\{\phi
_{k},\theta _{k}\},$ whose number corresponds to the number of
parameters determining the density matrix, equal to
$(2j+1)^{2}-1.$ But at the same time the dependence of the
tomogram $\mathcal{W}(m,U)$\ on the parameters of the unitary
matrix $U$ provides some advantage in considering the spins,
$j=0,1/2,1,\ldots ;$ and also the quantum states of several spins
in an unified approach. For two qudits (spin $j_{1}$ and $j_{2}$)
the tomogram of the quantum state with the
$(2j_{1}+1)(2j_{2}+1)\times (2j_{1}+1)(2j_{2}+1)$ density matrix \
$\rho $ is the normalized joint probability distribution
\begin{equation}
\mathcal{W}(m_{1},m_{2},U)=\left\langle m_{1}m_{2}\left\vert U\rho
U^{\dagger }\right\vert m_{1}m_{2}\right\rangle
\end{equation}
of two random spin projections $m_{1}=-j_{1},-j_{1}+1,\ldots
,j_{1}-1,j_{1}$ and $m_{2}=-j_{2},-j_{2}+1,\ldots ,j_{2}-1,j_{2}$
onto the corresponding directions determined by two pairs of Euler
angles, $\phi _{1},\theta _{1}$ and $\phi _{2},\theta _{2}.$ The
information contained in the tomogram with a dependence on the
matrix $U$ of such a form is sufficient to reconstruct the density
matrix $\rho .$ But we define the tomogram by Eq.(\ref{tomdef}) to
use the redundant information on the quantum state of bipartite
systems in studying the entanglement properties of the system
states. We remark that in Eq.(\ref{tomdef}) we could also use the
full unitary group instead of $SU(2)$ and this we will do
sometimes in the following.

The tomographic probability distribution of a qudit state
$\mathcal{W}(m,U)$ can be considered as a column vector
$\vec{\mathcal{W}}(U)$ with components
\begin{equation}
\mathcal{W}_{1}(U)=\mathcal{W}(j,U),\mathcal{W}_{2}(U)=\mathcal{W}%
(j-1,U),\ldots ,\mathcal{W}_{2j+1}(U)=\mathcal{W}(-j,U).
\end{equation}
Since all the components are nonnegative and the normalization condition (%
\ref{normcond}) holds, from a geometrical point of view the components $\{%
\mathcal{W}_{k}\}$ of the tomographic probability vector determine
the coordinates $\{x_{k}\}$ of points belonging to a simplex. For
a qubit such a simplex is the segment $\{x_{1}+x_{2}=1;0\leq
x_{1},x_{2}\leq 1\}$ in the
plane $x_{1},x_{2}.$ For a generic qudit the simplex is a polyhedron in a $%
(2j+1)-$dimensional space determined by equations:
\begin{equation}
\sum\limits_{k=1}^{2j+1}x_{k}=1;~0\leq x_{1},x_{2},\ldots
,x_{2j+1}\leq 1\ .
\end{equation}
Thus, the spin tomogram is a function of a unitary group element
$U$ with values in a simplex. The linear maps of probability
vectors
\begin{equation}
\vec{\mathcal{W}}^{\prime }(U)=M\vec{\mathcal{W}}(U)
\end{equation}
are expressed in terms of matrices ${M}$ which are known to form
the semigroup of stochastic matrices. A stochastic matrix is a
matrix with nonnegative entries such that the sum of the elements
of each column is one. If in addition the sum of the elements of
each row is one, the matrix is bistochastic. Also the bistochastic
matrices form a semigroup. In the next section we study the
properties of the tomograms and stochastic and bistochastic maps
on the example of qutrit states.

\section{Qutrit density matrix in tomographic probability representation}

To present the semigroup approach to describe generic qudit states
we start from qutrit states (spin-1 states). The density matrix of
a qutrit state is a nonnegative hermitian $3\times 3$ matrix with
unit trace and it can be presented in the product form
\begin{equation}
\rho =U_{0}\tilde{\rho}U_{0}^{\dagger }  \label{unop}
\end{equation}
where the $3\times 3$ matrix $\tilde{\rho}$ is diagonal:
\begin{equation}
\tilde{\rho}=\left(
\begin{array}{ccc}
\tilde{\rho}_{1} & 0 & 0 \\
0 & \tilde{\rho}_{2} & 0 \\
0 & 0 & \tilde{\rho}_{3}
\end{array}
\right)
\end{equation}
with nonnegative eigenvalues $\tilde{\rho}_{k},k=1,2,3,$
satisfying the normalization condition
\begin{equation}
\sum\limits_{k=1}^{3}\tilde{\rho}_{k}=1\ .  \label{simplex1}
\end{equation}
The columns of the $3\times 3$ unitary matrix $U_{0}$ are the
components of the normalized eigenvectors of the density matrix
$\rho $, i.e.
\begin{equation*}
U_{0}=\left(
\begin{array}{ccc}
u_{11} & u_{12} & u_{13} \\
u_{21} & u_{22} & u_{23} \\
u_{31} & u_{32} & u_{33}
\end{array}
\right) \equiv \left\vert \left\vert \vec{u}_{1},\vec{u}_{2},\vec{u}%
_{3}\right\vert \right\vert
\end{equation*}
such that
\begin{equation}
\rho \vec{u}_{1}=\tilde{\rho}_{1}\vec{u}_{1}\ ,\ \rho \vec{u}_{2}=\tilde{\rho%
}_{2}\vec{u}_{2}\ ,\ \rho \vec{u}_{3}=\tilde{\rho}_{3}\vec{u}_{3}\
.
\end{equation}
The eigenvalues $\tilde{\rho}_{k}$ can be considered as the
components of a probability vector
\begin{equation}
\vec{{\tilde{\rho}}}=\left[
\begin{array}{c}
\tilde{\rho}_{1} \\
\tilde{\rho}_{2} \\
\tilde{\rho}_{3}
\end{array}
\right]
\end{equation}
corresponding to a point on the triangle with vertices in
$(1,0,0),(0,1,0),(0,0,1),$ which is the two dimensional simplex
given by Eq.(\ref{simplex1}) with the constraints $0\leq
\tilde{\rho}_{k}\leq 1,\ k=1,2,3.$ The triangle (qutrit simplex)
is embedded
in the plane determined by Eq.(\ref{simplex1}), where the coordinates $%
\tilde{\rho}_{k}$ are arbitrary real numbers. The linear maps of
the qutrit probability vector $\vec{{\tilde{\rho}}}$ are
determined by the $3\times 3$ stochastic matrices $M,$ i.e.
\begin{equation}
\vec{{\tilde{\rho}}}\longrightarrow \vec{{\tilde{\rho}}}^{\prime }=M\vec{{%
\tilde{\rho}}}
\end{equation}
where the matrix $M$ has real nonnegative entries and the sum of
the elements of each column is unity. The product of two
stochastic matrices is again a stochastic matrix. Thus stochastic
matrices form a semigroup. Also, any convex sum of stochastic
matrices is a stochastic matrix. But one can check that the
inverse matrix $M^{-1}$ if it exists is not stochastic because
some entries are negative numbers. Thus the triangle (qutrit
simplex) is invariant under the action of a linear stochastic map
and the point on the plane (\ref{simplex1}) given by the vector $M\vec{{%
\tilde{\rho}}}$ belongs to the same triangle.

The tomogram $\mathcal{W}(m,U)$ of the qutrit state with density matrix $%
\rho $ given by Eq.(\ref{unop}) is the probability vector with
components
\begin{eqnarray}
\mathcal{W}_{1}(U) &=&\mathcal{W}(+1,U)=\left\langle 1\left\vert UU_{0}%
\tilde{\rho}U_{0}^{\dagger }U^{\dagger }\right\vert 1\right\rangle \ , \\
\mathcal{W}_{2}(U) &=&\mathcal{W}(0,U)=\left\langle 0\left\vert UU_{0}\tilde{%
\rho}U_{0}^{\dagger }U^{\dagger }\right\vert 0\right\rangle \ ,  \notag \\
\mathcal{W}_{3}(U) &=&\mathcal{W}(-1,U)=\left\langle -1\left\vert UU_{0}%
\tilde{\rho}U_{0}^{\dagger }U^{\dagger }\right\vert
-1\right\rangle \ . \notag
\end{eqnarray}
Direct calculation shows that the above formulae can be written in
the following form:
\begin{equation}
\mathcal{W}_{k}(U)=\sum\limits_{h=1}^{3}\left\vert \left(
UU_{0}\right) _{kh}\right\vert ^{2}\tilde{\rho}_{h}
\end{equation}
or in vector form:
\begin{equation}
\vec{\mathcal{W}}(U)=M\vec{{\tilde{\rho}}}  \label{bis1}
\end{equation}
where the elements of the matrix $M$ read
\begin{equation}
M_{kh}=\left\vert \left( UU_{0}\right) _{kh}\right\vert ^{2}\ .
\label{bis2}
\end{equation}
As a product of unitary matrices, the matrix $UU_{0}$ is unitary
and the sum of the (nonnegative) elements of each column and each
row of $M$ is one. So, $M$ is orthostochastic, a particular
bistochastic matrix. The result of Eq.s(\ref{bis1}),(\ref{bis2})
means that the qutrit state is determined by a bistochastic map
acting on the probability $3-$vector whose components are the
eigenvalues of the density matrix. The bistochastic matrix $M$ of
the map has elements which are the square moduli of the elements
of a unitary matrix. In turn, the unitary matrix is the product of
two unitary matrices, one rotating the basis in the space of spin
states and the other one having columns formed by the eigenvectors
of the density matrix. From a geometrical point of view, this
means that qutrit states are the orbit of the unitary group acting
on the points of the triangle (qutrit simplex), but the action of
the unitary group is made \textit{via} the action of an
orthostochastic map.

In the case of two qubits that construction yields the description
of the quantum states of the corresponding two spin$-1/2$ system
by the tomographic probability $4-$vector\ $\vec{\mathcal{W}}(U),$
where the the unitary matrix $U$ belongs to the group $U(4).$ The spin tomographic probability vector $\vec{\mathcal{W}}%
(U)$\ has the following components
\begin{eqnarray}
\mathcal{W}_{1}(U)
&=&\mathcal{W}(\frac{1}{2},\frac{1}{2},U)=\left\langle
\frac{1}{2}\frac{1}{2}\left\vert UU_{0}\tilde{\rho}U_{0}^{\dagger
}U^{\dagger }\right\vert \frac{1}{2}\frac{1}{2}\right\rangle \ , \\
\mathcal{W}_{2}(U)
&=&\mathcal{W}(\frac{1}{2},-\frac{1}{2},U)=\left\langle
\frac{1}{2}\frac{-1}{2}\left\vert UU_{0}\tilde{\rho}U_{0}^{\dagger
}U^{\dagger }\right\vert \frac{1}{2}\frac{-1}{2}\right\rangle \ ,  \notag \\
\mathcal{W}_{3}(U)
&=&\mathcal{W}(-\frac{1}{2},\frac{1}{2},U)=\left\langle
\frac{-1}{2}\frac{1}{2}\left\vert UU_{0}\tilde{\rho}U_{0}^{\dagger
}U^{\dagger }\right\vert \frac{-1}{2}\frac{1}{2}\right\rangle \ ,  \notag \\
\mathcal{W}_{4}(U)
&=&\mathcal{W}(-\frac{1}{2},-\frac{1}{2},U)=\left\langle
\frac{-1}{2}\frac{-1}{2}\left\vert
UU_{0}\tilde{\rho}U_{0}^{\dagger }U^{\dagger }\right\vert
\frac{-1}{2}\frac{-1}{2}\right\rangle \ .  \notag
\end{eqnarray}
Here the columns of the unitary $4\times 4-$matrix $U_{0}$ are the
eigenvectors and $\tilde{\rho}$ is the diagonal form of the
density matrix $\rho $ of the two qubits state. One can easily
check that the tomogram $\vec{\mathcal{W}}(U)$ can be written in
the form of the Eq.s(\ref{bis1}),(\ref{bis2}) if the eigenvalues
of $\rho$ are organized as a $4-$vector $\vec{{\tilde{\rho}}}$.
This vector belongs to a $3-$dimensional simplex, a polyhedron, in
a $4-$dimensional space. The points in the simplex belong to the
orbit of the group $U(4)$ in its Cartan subalgebra whose elements
are labelled by nonnegative numbers.

\section{The relation to general linear and inhomogeneous general linear
groups}

In this section we find a relation of stochastic and bistochastic
maps of the tomographic probability vectors to Lie groups. If we
restrict to invertible stochastic and bistochastic matrices and
leave out the nonnegativity of their entries, we obtain a group
since the invertible matrices have inverse matrices of the same
kind. To prove this let us formulate the stochasticity property in
terms of stability of a vector $\vec{e}_{0}$ with all its $n$
components equal to $1$. The (column) stochasticity property of an
$n\times n-$matrix $M$ results by the requirement:
\begin{equation}
e_{0}^{\mathrm{T}}M=e_{0}^{\mathrm{T}}  \label{stoc}
\end{equation}
where $e_{0}^{\mathrm{T}}=(1,1,\ldots ,1)$ is the transpose of $\vec{e}%
_{0} $. This requirement is equivalent to demand that the sum of
the elements of any column of $M$ is $1$. The identity matrix $I$
obviously satisfies the above equation. Since $I=MM^{-1}$ one has
\begin{equation}
e_{0}^{\mathrm{T}}=e_{0}^{\mathrm{T}}MM^{-1}=e_{0}^{\mathrm{T}}M^{-1}
\end{equation}
and this proves that Eq.(\ref{stoc}) is satisfied also by the
inverse matrix $M^{-1}$, when it exists.

The stochastic matrix $M$ is bistochastic if in addition to
Eq.(\ref{stoc}) it satisfies the condition
\begin{equation}
M\vec{e}_{0}=\vec{e}_{0}\ .  \label{bistoc}
\end{equation}
The same previous argument shows that also the inverse matrix
$M^{-1}$ satisfies Eq.(\ref{bistoc}), when it exists.

The invertible matrices $M$ satisfying equations (\ref{stoc}) and
(\ref
{bistoc})\ with real entries of arbitrary sign form a Lie group $\mathcal{G}%
_{BS}$. The subset of $\mathcal{G}_{BS}$ of invertible matrices
$M$ with real nonnegative entries is an open dense subsemigroup of
the bistochastic matrices semigroup.

Analogously, the invertible matrices $M$ satisfying equation
(\ref{stoc}) \ with real entries of arbitrary sign form a Lie
group $\mathcal{G}_{S}$.The subset of $\mathcal{G}_{S}$ of
invertible stochastic matrices is an open dense subsemigroup of
the stochastic matrices semigroup.

It is obvious that one can also consider complex matrices, rather
than real ones, satisfying the above equations. Then one obtains
other Lie groups.

It is clear from Eq.(\ref{bistoc}) that the Lie group $\mathcal{G}_{BS}$ is $%
GL(n-1,\mathbb{R})$ for the $n-$dimensional case, and
$GL(2,\mathbb{R})$ for the qutrit case.

Let us discuss the qutrit case in detail to recognize $%
\mathcal{G}_{BS}$ as $GL(2,\mathbb{R}).$ We choose a rotation
$\mathcal{O,}$
$\mathcal{OO}^{\mathrm{T}}=\mathcal{O}^{\mathrm{T}}\mathcal{O}=I$
, in the $3-$dimensional space such that
\begin{equation}
\frac{1}{\sqrt{3}}\mathcal{O}\left[
\begin{array}{c}
1 \\
1 \\
1
\end{array}
\right] =\left[
\begin{array}{c}
0 \\
0 \\
1
\end{array}
\right] \ .
\end{equation}
For instance, we choose:
\begin{equation}
\mathcal{O}=\left(
\begin{array}{ccc}
\frac{1}{\sqrt{2}} & \frac{-1}{\sqrt{2}} & 0 \\
\frac{1}{\sqrt{6}} & \frac{1}{\sqrt{6}} & \frac{-2}{\sqrt{6}} \\
\frac{1}{\sqrt{3}} & \frac{1}{\sqrt{3}} & \frac{1}{\sqrt{3}}
\end{array}
\right) \ .
\end{equation}
Let us take the bistochastic matrix $M$ of the form
\begin{equation}
M=\left(
\begin{array}{ccc}
x_{1} & y_{1} & z_{1} \\
x_{2} & y_{2} & z_{2} \\
x_{3} & y_{3} & z_{3}
\end{array}
\right)  \label{matform}
\end{equation}
with
\begin{equation}
\sum\nolimits_{k}x_{k}=\sum\nolimits_{k}y_{k}=\sum\nolimits_{k}z_{k}=1\
, \label{stocast}
\end{equation}
and
\begin{equation}
x_{k}+y_{k}+z_{k}=1\ ;\ k=1,2,3\ .  \label{bistocast}
\end{equation}
One can check that
\begin{equation}
\tilde{M}=\mathcal{O}M\mathcal{O}^{\mathrm{T}}=\left(
\begin{array}{ccc}
a & b & 0 \\
c & d & 0 \\
0 & 0 & 1
\end{array}
\right)  \label{matrot}
\end{equation}
where
\begin{eqnarray}
a &=&\frac{1}{2}(x_{1}-x_{2}-y_{1}+y_{2})\ ; \\
b &=&\frac{3}{2\sqrt{2}}(z_{2}-z_{1})\ ;  \notag \\
c &=&\frac{3}{2\sqrt{2}}(y_{3}-x_{3})\ ;  \notag \\
d &=&\frac{1}{2}(2z_{3}-x_{3}-y_{3})\ .  \notag
\end{eqnarray}
The $4-$parameter group of Eq.(\ref{matrot}) is the noncompact Lie group $%
GL(2,\mathbb{R}).$ As the determinant of the orthogonal matrix $\mathcal{O}$%
 is nonzero, the set of matrices $\tilde{M}$ is isomorphic to the
group of bistochastic matrices $M.$

Besides, if one considers the stochastic matrices Lie group
$\mathcal{G}_{S}$, that is matrices $M$ of the form
(\ref{matform}) satisfying Eq.(\ref {stocast}) but not
Eq.(\ref{bistocast}), one can check that
\begin{equation}
\tilde{M}=\mathcal{O}M\mathcal{O}^{\mathrm{T}}=\left(
\begin{array}{ccc}
A & B & m \\
C & D & n \\
0 & 0 & 1
\end{array}
\right)  \label{inmatrot}
\end{equation}
where
\begin{eqnarray}
A &=&\frac{1}{2}(x_{1}-x_{2}-y_{1}+y_{2})\ ; \\
\ B &=&\frac{1}{2\sqrt{2}}(x_{1}+y_{1}-2z_{1}-x_{2}-y_{2}+2z_{2})\
;  \notag
\\
C &=&\frac{3}{2\sqrt{2}}(y_{3}-x_{3})\ ;  \notag \\
D &=&\frac{1}{2}(2z_{3}-x_{3}-y_{3})\ ;  \notag \\
m &=&\frac{1}{\sqrt{6}}(x_{1}+y_{1}+z_{1}-x_{2}-y_{2}-z_{2})\ ;  \notag \\
n &=&\frac{1}{\sqrt{2}}(1-x_{3}-y_{3}-z_{3}))\ .  \notag
\end{eqnarray}

or equivalently

\begin{eqnarray}
x_{1} &=&\frac{1}{6}\left( 3A+\sqrt{2}B+\sqrt{2}C+D+\sqrt{6}m+\sqrt{2}%
n+2\right) , \\
x_{2} &=&\frac{1}{6}\left( -3A-\sqrt{2}B+\sqrt{2}C+D-\sqrt{6}m+\sqrt{2}%
n+2\right) , \notag \\
y_{1} &=&\frac{1}{6}\left( -3A+\sqrt{2}B-\sqrt{2}C+D+\sqrt{6}m+\sqrt{2}%
n+2\right) , \notag \\
y_{2} &=&\frac{1}{6}\left( 3A-\sqrt{2}B-\sqrt{2}C+D-\sqrt{6}m+\sqrt{2}%
n+2\right) , \notag \\
z_{1} &=&\frac{1}{6}\left( -2\sqrt{2}B-2D+\sqrt{6}m+\sqrt{2}n+2\right) , \notag  \\
z_{2} &=&\frac{1}{6}\left(
2\sqrt{2}B-2D-\sqrt{6}m+\sqrt{2}n+2\right) . \notag
\end{eqnarray}
From previous formulae one can readily see that pure translations
( $A=D=1$ and $B=C=0$ ) and pure homogeneous linear
transformations $(m=n=0)$ do not correspond to any stochastic
matrix.

The group of matrices of Eq.(\ref{inmatrot}) is just
$IGL(2,\mathbb{R}),$
isomorphic to the direct product of dilation group times $ISL(2,\mathbb{R})$%
\ , which in this particular dimension is isomorphic with the
Poincar\'{e} group in two plus one dimensions. In this way we have
established that the stochastic invertible semigroup of $3\times
3-$matrices is in one-to-one correspondence with a subset of the
group $IGL(2,\mathbb{R})$.

Finally, it is clear that this argument holds true for arbitrary
qudits (arbitrary spin$-j$ states).

Now we are able to describe briefly the action of the stochastic
and bistochastic groups on the carrier space $\mathbb{R}^{n}$ of
their above representation. The fiducial vector
$e_{0}^{\mathrm{T}}$ determines an invariant foliation given by
\begin{equation}
e_{0}^{\mathrm{T}}(M\vec{v})=e_{0}^{\mathrm{T}}(\vec{v})=\mathrm{const}\
;\ \vec{v}\in \mathbb{R}^{n}
\end{equation}
so that the simplex of probability vectors belong to the leaf $e_{0}^{%
\mathrm{T}}(\vec{v})=1.$ Moreover that simplex is invariant under
the action of the whole semigroup of stochastic matrices, and we
recall that the orbit of the semigroup of stochastic or even
bistochastic matrices starting from a vertex is the whole simplex.
While the action of the stochastic semigroup on the simplex is
transitive, the orbit of the bisthocastic semigroup starting from
a point $\vec{v}=\left( v_{1},...,v_{n}\right) ^{\mathrm{T}}$ is
the convex hull of all the vectors $\vec{v}_{\pi }$ whose
components are some permutation $\left( v_{\pi _{1}},...,v_{\pi
_{n}}\right) $ of the components of the given vector $\vec{v}.$

In particular, any point of the simplex can be also reached from a
given vertex with an invertible stochastic matrix. The same holds
true for invertible bistochastic matrix, if some lower dimensional
set of points of the simplex is dropped, in any case the point
$\vec{e}_{0}/n$ \ cannot be reached.

From the point of view of matrix analysis (for a general reference
of the following discussion see, e.g., \cite{Horn}), a generic
stochastic matrix $M$ is a positive matrix, i.e. all its entries
are positive. Then, by Perron's theorem, we know that its spectral
radius, which is $1,$ is an algebraic simple eigenvalue of maximum
modulus (Perron root of the matrix), with an eigenvector $\vec{p}$
(Perron vector) which may be chosen to be positive, i.e. a
probability vector.
\begin{equation}
M\vec{p}=\vec{p}\ ;\ p_{i}>0\ \forall i=1,..,n;\sum\nolimits_{i}\
p_{i}=1
\end{equation}
The fiducial vector $\vec{e}_{0}$ is just the left Perron vector
of the stochastic matrix $M$ belonging to the same eigenvalue $1,$
and the following limit does exist:
\begin{equation}
\lim_{k\rightarrow \infty
}M^{k}=L:=\vec{p}e_{0}^{\mathrm{T}}=\left\vert \left\vert
\vec{p},\vec{p},\ldots ,\vec{p}\right\vert \right\vert .
\end{equation}
In other words, the matrix $M^{k}$ approaches a limit which is a
rank-one stochastic matrix whose columns are the Perron vector of
$M.$ As a result,
because $L\vec{v}=\vec{p}$ for any vector of the leaf $e_{0}^{\mathrm{T}}(%
\vec{v})=1,$ we have that
\begin{equation}
\lim_{k\rightarrow \infty }M^{k}\vec{v}=\vec{p}
\end{equation}
independently of the starting point $\vec{v}$. In particular, if
$M$ is bistochastic, $\ \vec{p}=\vec{e}_{0}/n.$

We do not insist on the possibility of characterizing that limit
in terms of a stochastic Markov process.

When the stochastic matrix $M$ is nonnegative, rather than
positive, again a nonnegative probability eigenvector $\vec{p}$\
of the eigenvalue $1$\ exists, and we recover the same limit $L$
if the spectral radius of $M$ is the only eigenvalue of maximum
modulus.

More generally, if the matrix $I-M+L$ is invertible, we get the same limit $%
L $ by a Cesaro summation:
\begin{equation}
\lim_{N\rightarrow \infty }\frac{1}{N}\sum\limits_{k=1}^{N}M^{k}=L:=\vec{p}%
e_{0}^{\mathrm{T}}=\left\vert \left\vert \vec{p},\vec{p},\ldots ,\vec{p}%
\right\vert \right\vert .
\end{equation}

That hypothesis is crucial as the following example shows:
consider the permutation bistochastic matrix
\begin{equation}
M=\left(
\begin{array}{ccc}
0 & 1 & 0 \\
1 & 0 & 0 \\
0 & 0 & 1
\end{array}
\right)
\end{equation}
it gives a Cesaro limit
\begin{equation*}
\lim_{N\rightarrow \infty
}\frac{1}{N}\sum\limits_{k=1}^{N}M^{k}=\left(
\begin{array}{ccc}
\frac{1}{2} & \frac{1}{2} & 0 \\
\frac{1}{2} & \frac{1}{2} & 0 \\
0 & 0 & 1
\end{array}
\right)
\end{equation*}
which does not have the form of $L$. In any case the columns of
the Cesaro
limit, when it exists, are eigenvectors of $M$ belonging to the eigenvalue $%
1.$

\section{Positive maps}

Having discussed the transformations of probability vectors we are
now in the position of describing transformations of density
states by means of the tomographic representation of them.

The problem of constructing positive maps has a long history and
the dynamical maps providing a density matrix which could appear
in a process of quantum evolution was studied in \cite{Sudar61},
see also \cite{Kossak,Yamiolk}.

For finite level systems the space of states is a convex body in
the dual
space of the infinitesimal generators of the unitary group, say $u^{\ast }(%
\mathcal{H}).$ For closed quantum systems the dynamics is
conventionally represented by transformations associated with
one-parameter groups of unitary transformations in Hilbert spaces,
this formalism however is not appropriate to deal with
irreversible behaviours like decay of metastable particles,
approach to thermodynamic equilibrium and more generally to deal
with decoherence phaenomena.

For finite level quantum systems, probability measures are fully
described by probability vectors, say $n-$vectors $\vec{v}$ with
nonnegative components whose sum is $1$:
\begin{equation}
\left\{ \vec{v}:\ v_{m}\geq 0\ \forall m=1,\ldots ,n;\
\sum\nolimits_{m}v_{m}=1\right\} \ .
\end{equation}
For those systems, states are identified with $n\times n-$density matrices $%
\rho $ which are positive semi-definite and satisfy
\begin{equation}
\rho ^{\dagger }=\rho ,\ \mathrm{Tr}\rho =1
\end{equation}

Dynamical maps would assign to each $\rho $ another density state $\rho(t)=%
\mathbb{M}(t) \rho_{0},$ with $\mathbb{M}$ some $n^{2}\times
n^{2}-$matrix satisfying appropriate constraints to guarantee that
$\rho(t)$ is still a density state. A dynamical map on states
induces a dynamical map on probability vectors which may be
described by stochastic matrices. These maps, on states are called
quantum stochastic maps.

Our aim is to relate stochastic maps acting on probability vectors
to quantum stochastic maps acting on density states, by using the
tomographic setting. This approach is different from the standard
one\cite{SUD,Aso,Go} which describes quantum stochastic maps as
projection of isometries.

We may parametrize density states by a pair
\begin{equation}
\rho \rightarrow (U_{0},\vec{{\tilde{\rho}}})  \label{den map}
\end{equation}
where $U_{0},$ as a unitary matrix, provides us with the eigenvectors of $%
\rho $ while the components of $\vec{{\tilde{\rho}}}$ are the
corresponding eigenvalues. Symbolically we could write
\begin{equation}
\rho U_{0}=\vec{{\tilde{\rho}}}U_{0}  \label{ord}
\end{equation}
meaning that $\rho $ acting \ on the $k-$th column of $U_{0}$ will
provide the same column multiplied by the corresponding eigenvalue
$\rho _{k}.$ Equivalently, the columns of $U_{0}$ are an
orthonormal frame in the Hilbert
space and therefore define a family of orthonormal projectors \ while $\vec{{%
\tilde{\rho}}}$ gives the weights to attribute to each
corresponding rank-one projector of the decomposition of $\rho .$

It should be clear from this description that $U_{0}$ is determined up to $%
(U(1))^{n},$%
\begin{equation}
\rho =\sum_{m=1}^{n}\rho _{m}\left\vert e_{m}\right\rangle
\left\langle e_{m}\right\vert
\end{equation}
in a given chosen basis of orthonormal vectors. However if the
eigenvalues are degenerate, the ambiguity increases, replacing
each $U(1)$ subgroup with a $U(k)$ subgroup depending on the
degeneracy of each eigenvalue. Thus, in order to define
unambiguously the density matrix map of Eq.(\ref{den map}) we have
to choose a \textquotedblleft gauge\textquotedblright\ by fixing
the
phase factors of $U_{0}$ and an ordering of both the components of $\vec{{%
\tilde{\rho}}}$ \ and the columns of $U_{0}$ so that
Eq.(\ref{ord}) holds true.

This particular parametrization allows to deal more easily with
quantum stochastic maps $\rho \rightarrow \rho ^{\prime }$
parametrized in terms of unitary maps and stochastic maps,
providing therefore a different parametrization with respect to
those normally used in the literature. The density matrix $\rho $
is mapped onto another density matrix $\rho ^{\prime } $ in such a
way that the map is convex linear. This means that if two density
states are mapped onto other density states
\begin{equation}
\rho _{1}\rightarrow \rho _{1}^{\prime }\ ;\ \rho _{2}\rightarrow
\rho _{2}^{\prime }\ ,
\end{equation}
any their convex sum is mapped onto the same convex sum of the
images
\begin{equation}
\lambda _{1}\rho _{1}+\lambda _{2}\rho _{2}\rightarrow \lambda
_{1}\rho _{1}^{\prime }+\lambda _{2}\rho _{2}^{\prime }\ .
\end{equation}

In the tomographic framework one has the tomographic map of the
density matrix $\rho _{1}$\ onto the probability vector of the
qudit state
\begin{equation}
\rho _{1}\rightarrow \vec{\mathcal{W}}_{1}(U)=\left| UU_{01}\right| ^{2}\vec{%
{\tilde{\rho}}}_{1}
\end{equation}
and of the density density matrix $\rho _{2}$\ onto another
probability vector of the qudit state
\begin{equation}
\rho _{2}\rightarrow \vec{\mathcal{W}}_{2}(U)=\left| UU_{02}\right| ^{2}\vec{%
{\tilde{\rho}}}_{2}
\end{equation}

We introduce positive maps of density matrices, parametrized by an
unitary matrix $V$ and a stochastic matrix $M$, using this
tomographic setting in the following way:
\begin{eqnarray}
\left\vert UU_{01}\right\vert ^{2}\vec{{\tilde{\rho}}}_{1}
&\rightarrow
&\left\vert UU_{01}^{\prime }\right\vert ^{2}\vec{{\tilde{\rho}}}%
_{1}^{\prime }=\left\vert UVU_{01}\right\vert
^{2}M\vec{{\tilde{\rho}}}_{1}\
, \\
\left\vert UU_{02}\right\vert ^{2}\vec{{\tilde{\rho}}}_{2}
&\rightarrow
&\left\vert UU_{02}^{\prime }\right\vert ^{2}\vec{{\tilde{\rho}}}%
_{2}^{\prime }=\left\vert UVU_{02}\right\vert
^{2}M\vec{{\tilde{\rho}}}_{2}\ .  \notag
\end{eqnarray}
Thus our map is equivalent to the map of the pair $(U_{0},\vec{{\tilde{\rho}}%
})$ onto another analogous pair $(U_{0}^{\prime },\vec{{\tilde{\rho}}}%
^{\prime })$ obtained by acting with an unitary matrix $V$ and \ a
stochastic matrix $M$ as:
\begin{equation}
(U_{0},\vec{{\tilde{\rho}}})\rightarrow (U_{0}^{\prime },\vec{{\tilde{\rho}}}%
^{\prime })=(VU_{0},M\vec{{\tilde{\rho}}})\ .
\end{equation}
Since any density matrix $\rho $ is completely determined by the
corresponding pair $(U_{0},\vec{{\tilde{\rho}}})$ and as the left
actions on the unitary group and on the simplex by stochastic maps
are transitive, we have described a class of positive maps which
act transitively on density matrices.

To illustrate that the above map is convex linear we discuss in
detail the simple qubit case. To do this first we find explicit
formulae for the maps of density matrix $\rho $ and pair
$(U_{0},\vec{{\tilde{\rho}}})$
\begin{equation}
\rho \rightarrow (U_{0},\vec{{\tilde{\rho}}})  \label{m1}
\end{equation}
and \textit{viceversa }
\begin{equation}
(U_{0},\vec{{\tilde{\rho}}})\rightarrow \rho .  \label{m2}
\end{equation}
As we have already remarked the map of Eq.(\ref{m1}) requires a
\textquotedblleft gauge\textquotedblright\ choice because we are
going from a three dimensional manifold to a five dimensional one.

Let the qubit density matrix be
\begin{equation}
\rho =
\begin{pmatrix}
\rho _{11} & \rho _{12} \\
\rho _{21} & \rho _{22}
\end{pmatrix}
\ ,\ 0\leq \mathrm{\det }\rho \leq \frac{1}{4}.  \label{m3}
\end{equation}
Then its eigenvalues read
\begin{eqnarray}
\tilde{\rho}_{1} &=&\frac{1}{2}+\frac{1}{2}\sqrt{1-4\mathrm{\det
}\rho }
\label{m4} \\
\tilde{\rho}_{2} &=&\frac{1}{2}-\frac{1}{2}\sqrt{1-4\mathrm{\det
}\rho }. \label{m5}
\end{eqnarray}
The eigenvector corresponding to $\tilde{\rho}_{1}$ is
\begin{equation}
\vec{u}_{1}=
\begin{pmatrix}
u_{11} \\
u_{21}
\end{pmatrix}
\label{m6}
\end{equation}
where
\begin{eqnarray}
u_{11} &=&\frac{2\rho _{12}}{1+\sqrt{1-4\mathrm{\det }\rho }-2\rho _{11}}%
y_{1}  \label{m7} \\
u_{21} &=&y_{1}=\left( \left\vert \frac{2\rho
_{12}}{1+\sqrt{1-4\mathrm{\det }\rho }-2\rho _{11}}\right\vert
^{2}+1\right) ^{-\frac{1}{2}}.  \label{m8}
\end{eqnarray}
The eigenvector corresponding to $\tilde{\rho}_{2}$ is
\begin{equation}
\vec{u}_{2}=
\begin{pmatrix}
u_{12} \\
u_{22}
\end{pmatrix}
\label{m9}
\end{equation}
where
\begin{eqnarray}
u_{12} &=&\frac{2\rho _{12}}{1-\sqrt{1-4\mathrm{\det }\rho }-2\rho _{11}}%
y_{2}  \label{m10} \\
u_{22} &=&y_{2}=\left( \left\vert \frac{2\rho
_{12}}{1-\sqrt{1-4\mathrm{\det }\rho }-2\rho _{11}}\right\vert
^{2}+1\right) ^{-\frac{1}{2}},  \label{m11}
\end{eqnarray}
and the phases have been chosen to be zero. Thus given the qubit matrix $%
\rho $ of Eq.(\ref{m3}) one has the corresponding pair of unitary
matrix
\begin{equation}
U_{0}=
\begin{pmatrix}
u_{11} & u_{12} \\
u_{21} & u_{22}
\end{pmatrix}
\label{m12}
\end{equation}
with components (\ref{m7}), (\ref{m8}), (\ref{m10}), (\ref{m11})
and the point on simplex given by the probability vector
\begin{equation}
\vec{{\tilde{\rho}}}=
\begin{pmatrix}
\tilde{\rho}_{1} \\
\tilde{\rho}_{2}
\end{pmatrix}
\label{m13}
\end{equation}
with components (\ref{m4}), (\ref{m5}). One can check that the
matrix $\rho $ of Eq.(\ref{m3}) has the representation
\begin{equation}
\rho =U_{0}
\begin{pmatrix}
\tilde{\rho}_{1} & 0 \\
0 & \tilde{\rho}_{2}
\end{pmatrix}
U_{0}^{\dagger }.  \label{m14}
\end{equation}

Let us consider the inverse problem. Namely, given a unitary
matrix $V_{0}$ of the form (\ref{m12}) and a probability vector
\begin{equation}
\vec{\chi}=
\begin{pmatrix}
\chi _{1} \\
\chi _{2}
\end{pmatrix}
,\chi _{1}+\chi _{2}=1  \label{m16}
\end{equation}
describing a point on the simplex, let us construct the matrix
\begin{equation}
\chi =V_{0}
\begin{pmatrix}
\chi _{1} & 0 \\
0 & \chi _{2}
\end{pmatrix}
V_{0}^{\dagger }=
\begin{pmatrix}
\chi _{11} & \chi _{12} \\
\chi _{21} & \chi _{22}
\end{pmatrix}
\label{m17}
\end{equation}
whose matrix elements explicitly are:
\begin{eqnarray}
\chi _{11} &=&\left\vert V_{11}\right\vert ^{2}\chi
_{1}+\left\vert
V_{12}\right\vert ^{2}\chi _{2}=1-\chi _{22}\ ;  \label{m18} \\
\chi _{12} &=&V_{11}V_{21}^{\ast }\chi _{1}+V_{12}V_{22}^{\ast
}\chi _{2}=\chi _{21}^{\ast }\ .  \label{m19}
\end{eqnarray}

Thus, given the pair $(V_{0},\vec{\chi}),$ one has a density
matrix $\chi $ with the previous matrix elements. One can see that
the map $\rho \rightarrow (U_{0},\vec{{\tilde{\rho}}})$ is
nonlinear. The formulae (\ref {m18}), (\ref{m19}) provide a
polynomial dependence of the elements of density matrix on unitary
matrix and probability vector components. The dependence of
density matrix elements on probability vector components is linear
and on unitary matrix elements $V_{jk}$ is quadratic. It is
obvious that the inverse transform $\rho \rightarrow
(U_{0},\vec{{\tilde{\rho}}})$
is also nonlinear as formulae (\ref{m4}), (\ref{m5}), (\ref{m7}), (\ref{m8}%
), (\ref{m10}), (\ref{m11}) show. Moreover formula (\ref{m18})
shows how two $U(1)$ elements disappear in defining $\chi _{jk}.$

Let us discuss now the representation of a convex sum of two
density matrices $\rho ^{(1)}$ and $\rho ^{(2)}$ given by
\begin{equation}
\rho =\lambda _{1}\rho ^{(1)}+\lambda _{2}\rho ^{(2)},\ \lambda
_{1}+\lambda _{2}=1,\ 0\leq \lambda _{j}\leq 1  \label{m20}
\end{equation}
The matrix elements of this matrix read
\begin{equation}
\rho _{jk}=\lambda _{1}\rho _{jk}^{(1)}+\lambda _{2}\rho
_{jk}^{(2)} \label{m21}
\end{equation}
The two eigenvectors of this matrix are
\begin{equation}
\tilde{\rho}_{1,2}=\frac{1}{2}\pm \frac{1}{2}\sqrt{1-4\lambda _{1}^{2}%
\mathrm{\det }\rho ^{(1)}-4\lambda _{2}^{2}\mathrm{\det }\rho
^{(2)}-4\lambda _{1}\lambda _{2}\gamma }  \label{m22}
\end{equation}
where
\begin{equation}
\gamma =\rho _{11}^{(1)}\rho _{22}^{(2)}+\rho _{11}^{(2)}\rho
_{22}^{(1)}-\rho _{12}^{(1)}\rho _{21}^{(2)}-\rho _{12}^{(2)}\rho
_{21}^{(1)} \label{m23}
\end{equation}
One can check the following equality of probability vectors:
\begin{equation}
\left\vert UU_{0}(\vec{\lambda})\right\vert ^{2}\vec{{\tilde{\rho}}}(\vec{%
\lambda})=\lambda _{1}\left\vert UU_{0}^{(1)}\right\vert ^{2}\vec{{\tilde{%
\rho}}}_{1}+\lambda _{2}\left\vert UU_{0}^{(2)}\right\vert ^{2}\vec{{\tilde{%
\rho}}}_{2}  \label{m24}
\end{equation}
where $\vec{{\tilde{\rho}}}(\vec{\lambda})$ is the probability
vector with
components given by Eq.s(\ref{m22}), (\ref{m23}). The unitary matrix $U_{0}(%
\vec{\lambda})$ is obtained by formula (\ref{m12}) with entries (\ref{m7}), (%
\ref{m8}), (\ref{m10}), (\ref{m11}) after the replacements
\begin{eqnarray}
&&\rho _{12}\rightarrow \lambda _{1}\rho _{12}^{(1)}+\lambda
_{2}\rho
_{12}^{(2)} \\
&&\rho _{11}\rightarrow \lambda _{1}\rho _{11}^{(1)}+\lambda
_{2}\rho
_{11}^{(2)} \notag \\
&&\mathrm{\det }\rho \rightarrow \lambda _{1}^{2}\mathrm{\det
}\rho ^{(1)}+\lambda _{2}^{2}\mathrm{\det }\rho ^{(2)}+\lambda
_{1}\lambda _{2}\gamma \notag
\end{eqnarray}
where $\gamma $ is given by Eq.(\ref{m23}). Thus we checked
directly the convex superposition formula for two tomographic
probability vectors corresponding to a convex mixture of two density matrix $%
\rho ^{(1)}$ and $\rho ^{(2)}$ for two qubits. From Eq.(\ref{m24})
one can readily see that
\begin{equation}
\vec{{\tilde{\rho}}}(\vec{\lambda})=\lambda _{1}\left\vert U_{0}^{-1}(\vec{%
\lambda})U_{0}^{(1)}\right\vert
^{2}\vec{{\tilde{\rho}}}_{1}+\lambda
_{2}\left\vert U_{0}^{-1}(\vec{\lambda})U_{0}^{(2)}\right\vert ^{2}\vec{{%
\tilde{\rho}}}_{2}.
\end{equation}
Thus we have illustrated in detail that the map of pairs $(U_{0},\vec{{%
\tilde{\rho}}})\rightarrow (VU_{0},M\vec{{\tilde{\rho}}})$ by
means of left action on unitary group and stochastic matrix map on
simplex is convex
linear. It is worth to note that the map $(U_{0},\vec{{\tilde{\rho}}}%
)\rightarrow (U_{0}^{\ast },\vec{{\tilde{\rho}}})$ , which is just
the transposition of the density matrix map, is not equivalent to
left multiplication by a matrix. So, we can extend the map using
automorphism groups of both the unitary matrices and simplex
points.

\section{Entanglement}

Stochastic maps may be used successfully to characterize the
entanglement of states for a composite physical system in the
tomographic scheme as the following example shows.

To study entangled states we consider two qubit states with state
vector
\begin{equation}
\left\vert \psi \right\rangle =\frac{1}{\sqrt{2}}\left( \left\vert +\frac{1}{%
2}\right\rangle \left\vert -\frac{1}{2}\right\rangle +\left\vert -\frac{1}{2}%
\right\rangle \left\vert +\frac{1}{2}\right\rangle \right) .
\label{eq30}
\end{equation}
The density matrix of this state reads
\begin{equation}
\rho =\left\vert \psi \right\rangle \left\langle \psi \right\vert =\frac{1}{2%
}
\begin{pmatrix}
0 & 0 & 0 & 0 \\
0 & 1 & 1 & 0 \\
0 & 1 & 1 & 0 \\
0 & 0 & 0 & 0
\end{pmatrix}
\ ,  \label{eq31}
\end{equation}
so its four eigenvalues yield the probability vector
\begin{equation}
\overrightarrow{\widetilde{\rho }}=
\begin{bmatrix}
0 \\
1 \\
0 \\
0
\end{bmatrix}
\ ,
\end{equation}
while the corresponding eigenvectors form $U_{0}:$%
\begin{equation}
U_{0}=
\begin{pmatrix}
1 & 0 & 0 & 0 \\
0 & \frac{1}{\sqrt{2}} & \frac{1}{\sqrt{2}} & 0 \\
0 & \frac{1}{\sqrt{2}} & -\frac{1}{\sqrt{2}} & 0 \\
0 & 0 & 0 & 1
\end{pmatrix}
\ .
\end{equation}
In view of this, the tomographic probability vector of the two
qubit state has the following form:
\begin{equation}
\mathcal{\vec{W}}(U)=\frac{1}{2}\left[
\begin{array}{c}
|u_{12}+u_{13}|^{2} \\
|u_{22}+u_{23}|^{2} \\
|u_{32}+u_{33}|^{2} \\
|u_{42}+u_{43}|^{2}
\end{array}
\right] \ ,
\end{equation}
where the $u$'s are the matrix elements of $U.$ One can recognize
the entanglement of the state with this tomogram, calculating the
stochastic
matrix $M$ whose columns are four probability vectors $\mathcal{\vec{W}}%
(U_{ab}),\mathcal{\vec{W}}(U_{ac}),\mathcal{\vec{W}}(U_{db}),\mathcal{\vec{W}%
}(U_{dc}),$ where each unitary matrix $U_{hk}$ is a tensor product of two $%
2\times 2$ unitary matrices $U_{h}\otimes U_{k}:$%
\begin{equation}
U_{hk}=U_{h}\otimes U_{k}\ ,\ (h=a,d\ ;\ k=b,c)\ .
\end{equation}
Eventually, the stochastic matrix $M$ has the form
\begin{equation}
M=\left(
\begin{array}{cccc}
x_{ab} & x_{ac} & x_{db} & x_{dc} \\
\frac{1}{2}-x_{ab} & \frac{1}{2}-x_{ac} & \frac{1}{2}-x_{db} & \frac{1}{2}%
-x_{dc} \\
\frac{1}{2}-x_{ab} & \frac{1}{2}-x_{ac} & \frac{1}{2}-x_{db} & \frac{1}{2}%
-x_{dc} \\
x_{ab} & x_{ac} & x_{db} & x_{dc}
\end{array}
\right) \ .
\end{equation}
where $x_{hk}$ is a trigonometric function of the Euler angles determining $%
U_{h},U_{k}$ .

To evaluate the Bell number $B$ satisfying Bell's
inequality\cite{Bell64,CHSH}
\begin{equation*}
B\leq 2
\end{equation*}
one has to calculate\cite{Lupo} a trace:
\begin{equation}
B=\mathrm{Tr}(ME)
\end{equation}
where $E$ is the following matrix:
\begin{equation}
E=
\begin{pmatrix}
1 & -1 & -1 & 1 \\
1 & -1 & -1 & 1 \\
1 & -1 & -1 & 1 \\
-1 & 1 & 1 & -1
\end{pmatrix}
\ .
\end{equation}
The elements of the matrix $M$ are functions of four directions.
The Cirelson bound\cite{Cirel} for the Bell number is $2\sqrt{2}$
and can be achieved in entangled states only. One can check that
the bound is achieved when
\begin{equation}
x_{ab}=x_{ac}=x_{db}=x\ ;\ x_{dc}=1-x\ .
\end{equation}
In that case one has
\begin{equation}
2\sqrt{2}=4(4x-1)\ ;\ x=\frac{2+\sqrt{2}}{8}\ .
\end{equation}
Thus the universal stochastic matrix corresponding to an
arbitrary, maximally entangled state of two qubit is
\begin{equation}
M=
\begin{pmatrix}
\frac{2+\sqrt{2}}{8} & \frac{2+\sqrt{2}}{8} & \frac{2+\sqrt{2}}{8} & \frac{2-%
\sqrt{2}}{8} \\
\frac{2-\sqrt{2}}{8} & \frac{2-\sqrt{2}}{8} & \frac{2-\sqrt{2}}{8} & \frac{2+%
\sqrt{2}}{8} \\
\frac{2-\sqrt{2}}{8} & \frac{2-\sqrt{2}}{8} & \frac{2-\sqrt{2}}{8} & \frac{2+%
\sqrt{2}}{8} \\
\frac{2+\sqrt{2}}{8} & \frac{2+\sqrt{2}}{8} & \frac{2+\sqrt{2}}{8} & \frac{2-%
\sqrt{2}}{8}
\end{pmatrix}
\ .
\end{equation}

Thus we have clarified what is the relation of stochastic matrices
with Bell inequality violation for entangled states of two qubits.

\section{Conclusions}

We point out the main results of this paper. We have established a
connection of spin tomograms with stochastic maps acting on a
simplex and
unitary group elements. We have shown that stochastic and bistochastic $%
n\times n$ matrices have a dense intersection with the Lie groups $IGL(n-1,%
\mathbb{R})$ and $GL(n-1,\mathbb{R})$ respectively. We have
constructed positive maps of density states as maps determined by
pairs of unitary matrices and stochastic matrices. We have
demonstrated that for entangled two qubit states the Cirelson
bound for the Bell number is associated with a universal
stochastic matrix.

To conclude, we observe that in description of quantum states an
important role is played by unitary irreducible representations of
Lie groups. One can see that starting with unitary representations
of the $GL(2,\mathbb{R})$ group which are infinite-dimensional and
restricting the representations to the semigroup subset one
obtains unitary representations of invertible bistochastic matrix
semigroup. Analogously the unitary representations of the
stochastic semigroup can be obtained by restricting to the
semigroup subset the infinite-dimensional unitary representations
of the inhomogeneous linear group (either real or complex). The
problem of irreducibility or reducibility of the semigroup
representations has to be analyzed further.

Finally, we hope to clarify the relations between the constructed
positive map and other existing approaches in a future work.
\newline

\noindent\textbf{Acknowledgments}

V.I.M. thanks the University "Federico II" of Naples for kind
hospitality and the Russian Foundation for Basic Research for
partial support under Project No.~07-02-00598.

\end{document}